\documentclass[11pt]{article}%
\pdfoutput=1
\usepackage{amssymb}
\usepackage[left=1.1in,right=1.1in,top=1in,bottom=1in]{geometry}
\usepackage{amsmath}
\usepackage{amsfonts}
\usepackage{graphicx}
\usepackage{float}
\usepackage{endnotes}
\usepackage{epigraph}%
\providecommand{\U}[1]{\protect\rule{.1in}{.1in}}
\let\footnote=\endnote
\begin{document}

\title{Patterns in the Fabric of Nature}
\author{Steven Weinstein\thanks{sw@uwaterloo.ca; sweinstein@perimeterinstitute.ca}\\Perimeter Institute for Theoretical Physics, 31 Caroline St N, Waterloo, ON
\ N2L 2Y5\\U. Waterloo Dept. of Philosophy, 200 University Ave West, Waterloo, ON\ N2L 3G1}
\date{}
\maketitle

\begin{abstract}
From classical mechanics to quantum field theory, the physical facts at one
point in space are held to be independent of those at other points in space.
\ I propose that we can usefully challenge this orthodoxy in order to explain
otherwise puzzling correlations at both cosmological and microscopic scales.

\epigraph{``Nature uses only the longest threads to weave her patterns, so that each small piece of her fabric reveals the organization of the entire tapestry.''}{\em{The Character of Physical Law}, Richard Feynman}

\end{abstract}

\section{Introduction}

Despite radical differences in their conceptions of space, time, and the
nature of matter, all of the physical theories we presently use ---
non-relativistic and relativistic, classical and quantum --- share one
assumption:\ the features of the world at distinct points in space are
understood to be independent. Particles may exist anywhere, independent of the
location or velocity of other particles. Classical fields may take on any
value at a given point, constrained only by local constraints like Gauss's
law. \ Quantum field theories incorporate the same independence in their
demand that field operators at distinct points in space commute with one another.

The independence of physical properties at distinct points is a theoretical
assumption, albeit one that is grounded in our everyday experience.\ We appear
to be able to manipulate the contents of a given region of space unrestricted
by the contents of other regions. We can arrange the desk in our office
without concern for the location of the couch at home in our living room.

Yet there are realms of physical theory, more remote from everyday experience
and physical manipulation yet accessible to observation, in which there appear
to be striking correlations between the values of physical properties at
different points in space. Quantum theory predicts (and experiment confirms)
the existence of strongly correlated measurement outcomes apparently
inexplicable by classical means. I refer, of course, to the measurements of
the properties of pairs of particles originally envisioned by Einstein,
Podolsky and Rosen (EPR) \cite{EPR35}, measurements that suggested to EPR\ the
incompleteness of the theory. Bell (1964) showed that no theory satisfying two
seemingly natural conditions could possibly account for these correlations.
Quantum mechanics itself violates one of the conditions, known as \emph{Bell
locality}, \emph{strong locality} or \emph{factorizability}, leading to the
measurement paradoxes that so troubled Einstein and his collaborators. The
other condition,\emph{ statistical independence}, has only rarely been
questioned. It is tantamount to the rejection of the assumption of local
degrees of freedom, or the existence of nonlocal constraints.

On a completely different scale, the electromagnetic radiation that pervaded
the early universe -- the remnants of which form the cosmic microwave
background -- appears to have been extraordinarily homogeneous.\ It is
strikingly uniform, yet the theories that describe the early universe --
classical electrodynamics (for the radiation) and general relativity (for the
expanding spacetime the radiation fills) -- do not stipulate any sort of
restrictions or correlations that would go anywhere near explaining this. To
the extent that they have been explained at all, it has been through the
postulation of an as-yet unobserved field known as the inflaton field.

What I want to do here is raise the possibility that there is a more
fundamental theory possessing nonlocal constraints that underlies our current
theories. Such a theory might account for the mysterious nonlocal effects
currently described, but not explained, by quantum mechanics, and might
additionally reduce the extent to which cosmological models depend on finely
tuned initial data to explain the large scale correlations we observe. The
assumption that spatially separated physical systems are entirely uncorrelated
is a parochial assumption borne of our experience with the everyday objects
described by classical mechanics. Why not suppose that at certain scales or
certain epochs, this independence emerges from what is otherwise a highly
structured, nonlocally correlated microphysics?

\section{Nonlocal constraints}

All physical theories in current use assume that the properties of physical
systems at different points in space are independent. Correlations may emerge
dynamically -- many liquids crystallize and develop a preferred orientation
when cooled, for example -- but the fundamental theories permit any
configuration as an initial condition.

For example, consider the straightforward and simple theory of the free
massless scalar field $\phi(\vec{x})$. A scalar field is simply an assignment
of a single number (a \textquotedblleft scalar\textquotedblright\ rather than
a vector) to every point in space and time. The evolution of the field is
given by the well-known wave equation
\[
\frac{\partial^{2}\phi(\vec{x},t)}{\partial t^{2}}=c^{2}\nabla^{2}\phi(\vec
{x},t)\text{ ,}%
\]
in conjunction with initial data $\phi(\vec{x})$ and $\partial\phi(\vec
{x})/\partial t$ giving the value of the field and its rate of change at some
initial time. This initial data can be specified arbitrarily --- it is unconstrained.

A more realistic field theory is the classical electrodynamics of Maxwell,
which \emph{does }feature constraints. In Maxwell's theory, we have a pair of
coevolving fields, the electric field $\overrightarrow{E}$ and the magnetic
field $\overrightarrow{B}$. The fields are described by vectors at each point
rather than scalars.\ The significant difference between the electromagnetic
field and the free scalar field is that the electric and magnetic fields may
not be specified arbitrarily. They are subject to constraints $\nabla
\cdot\overrightarrow{E}(\vec{x})=4\pi\rho(\vec{x})$ and $\nabla\cdot\vec
{B}(\vec{x})=0$ which hold at every point $\vec{x}$ in space. The divergence
of the electric field at any given point with coordinates must be equal to a
multiple of the charge density $\rho(\vec{x})$ at that point, and the
divergence of the magnetic field must be zero. The divergence is a measure of
the outflow of the field in the neighborhood of a point, and the two
constraints tell us respectively that any such outflow of the electric field
is due to the presence of a charge at that point acting as a source, while the
magnetic field can have no sources (there are no magnetic charges). These
constraints are \emph{local} in that they provide a constraint on values of
the field at each point that does not depend on values of the field or the
charge distribution at other points.

What would a nonlocal constraint look like? Here's a candidate: $\nabla
\cdot\overrightarrow{E}(\vec{x})=4\pi\rho(\vec{x}-(1,1,1)).$ This says that
the divergence of the electric field at one point is equal to a constant times
the charge density at a point which is one unit away in each of the $x,y$ and
$z$ directions. But this constraint is hardly worthy of the name, since it
only holds at a single time: unlike the constraint $\nabla\cdot E(\vec
{x})=4\pi\rho(\vec{x})$ , it is not preserved by the equations of motion
(Maxwell's equations for the field and the Lorentz force law for the charge
distribution). I.e., it may hold at one time, but will not continue to hold as
the field evolves. Since it is not preserved, it does not hold at arbitrary
moments of time, hence it is not a true regularity or law.

Let's return to simple mechanics for an example of a true nonlocal constraint,
one that is conserved in time. The particles are characterized by their
positions and their momenta. The constraint we will impose is that the total
momentum (the sum of the momenta of each of the particles) is zero. This is a
constraint because we cannot specify the momentum freely for each particle; if
we know the momentum of all but one of the particles, the momentum of the
other particle is fixed.\ It is nonlocal, because the momentum of that
particle depends on the momenta of particles some distance away.\ Unlike the
first nonlocal constraint we considered, it is conserved, since total momentum
is a conserved quantity in particle mechanics.\ But it is not a particularly
interesting constraint, because all but one of the momenta may be freely
specified. Whereas the two constraints in electromagnetism reduce the number
of degrees of freedom from six to four at each point in space (so that there
are only two-thirds the number of degrees of freedom), this constraint only
reduces the total number of degrees of freedom by one.

\begin{figure}[ph]%
\centering
\includegraphics[
trim=0.000000in 0.000000in 0.001982in 0.000000in,
natheight=2.846100in,
natwidth=6.607200in,
height=1.5082in,
width=3.4627in
]%
{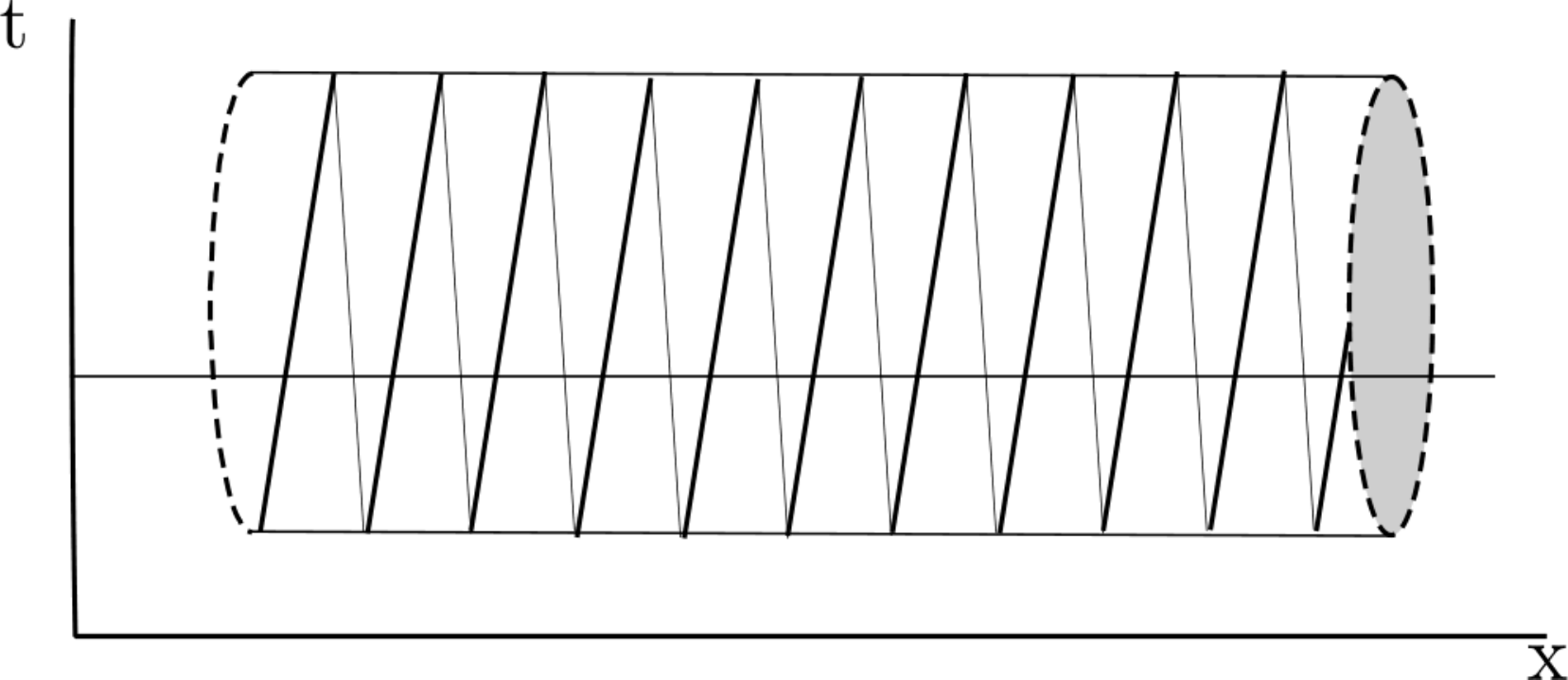}%
\caption{Timelike compactification}%
\label{CTC}%
\end{figure}

A\ more interesting nonlocal constraint may be obtained by considering once
more the wave equation, this time in one space dimension (for simplicity).
Suppose that the spacetime on which the field takes values is compactified in
the time direction, so that the entirety forms a cylinder (see Figure
\ref{CTC}).
%TCIMACRO{\FRAME{fphFU}{3.4627in}{1.5082in}{0pt}{\Qcb{Timelike
%compactification}}{\Qlb{CTC}}{CTCnew.pdf}%
%width 3.4627in;  height 1.5082in;  depth 0pt;  original-width 6.6072in;
%original-height 2.8461in;  cropleft "0";  croptop "1";  cropright "0.9997";
%cropbottom "0";  filename 'CTCnew.pdf';file-properties "XNPEU";}}}%
%BeginExpansion
%EndExpansion

\noindent The solutions must clearly be periodic, and this amounts to imposing
a nonlocal constraint. More specifically, whereas in the ordinary initial
value problem, initial data may be any smooth functions $\phi(x,0)$ and
$\phi_{t}(x,0)$ (where $\phi_{t}$ stands for $\partial\phi/\partial t$) , we
now require that $\phi(x,0)=\phi(x,T)$ and $\phi_{t}(x,0)=\phi_{t}(x,T)$,
where $T$ is the circumference of the cylinder. This is just to say that the
time evolution from $0$ to $T$ must return us to the same starting point. What
are the constraints, then, on this initial data? They are essentially those
data that can be written as sums of sine or cosine waves with wavelength
$\frac{T}{2\pi n}$ (for any integer value of $n$).\footnote{Solutions to the
wave equation can be written as sums of plane waves, with Fourier space
representation $\hat{\phi}(k,t)=\hat{F}(k)e^{-ikt}+\hat{G}(k)e^{ikt}$. Since
these plane waves must have period $T$ (in the preferred frame dictated by the
cylinder), we have a constraint $k=\frac{2\pi n}{T}$ (where $n$ is a positive
or negative integer), so that initial data are no longer arbitrary smooth
functions of $k$
\begin{align*}
\hat{\phi}(k,0) &  =\hat{F}(k)+\hat{G}(k)\\
\hat{\phi}_{t}(k,0) &  =-ik(\hat{F}(k)-\hat{G}(k))
\end{align*}
but are rather constrained by the requirement $k=\frac{2\pi n}{T}$. Thus the
initial data are the functions%
\begin{align*}
\phi(x,0) &  =\frac{1}{\sqrt{T}}%
%TCIMACRO{\dsum \limits_{n=-\infty}^{\infty}}%
%BeginExpansion
{\displaystyle\sum\limits_{n=-\infty}^{\infty}}
%EndExpansion
\hat{\phi}(\frac{2\pi n}{T},0)e^{i\frac{2\pi n}{T}x}dk\\
\phi_{t}(x,0) &  =\frac{1}{\sqrt{T}}%
%TCIMACRO{\dsum \limits_{n=-\infty}^{\infty}}%
%BeginExpansion
{\displaystyle\sum\limits_{n=-\infty}^{\infty}}
%EndExpansion
\hat{\phi}_{t}(\frac{2\pi n}{T},0)e^{i\frac{2\pi n}{T}x}dk
\end{align*}
i.e., they consist of arbitrary sums of plane waves with wave number
$k=\frac{2\pi n}{T}$, for any integer value of $n$.}

The restriction to a discrete (though infinite) set of plane waves means that
initial data do not have compact support; they are periodic. However, for
sufficiently large $T$ or sufficiently small $\Delta x$, the local physics is
indistinguishable from the local physics in ordinary spacetime. Only at
distance scales on the order of $T$ does the compact nature of the direction
become evident in the repetition of the spatial structure. Thus we have here
an example of a nonlocal constraint which might give the appearance of
unconstrained local degrees of freedom.

Now, this spacetime obviously has closed timelike curves, and it is
interesting to note that under such conditions, classical computers are as
powerful as quantum computers \cite{AW08}. Thus there is some reason to think
that a nonlocal constraints might allow one to mimic other quantum phenomena
using classical physics. In any event, we will now proceed to a discussion of
the way in which the presence of nonlocal constraints opens the door to a
little-explored loophole in Bell's theorem, in that their presence undermines
the \emph{statistical independence} assumption required for the proof of the theorem.

\section{Bell's theorem}

Einstein believed quantum theory to be an incomplete description of the world,
and he and his collaborators Podolsky and Rosen attempted to show this in
their 1935 paper \cite{EPR35}. The argument involves a pair of particles
specially prepared in an entangled state of position and
momentum.\footnote{The state used by EPR is an eigenstate of the operators
representing the sum of the momenta and the difference of the positions of the
two particles.} Quantum mechanics makes no definite predictions for the
position and momentum of each particle, but does make unequivocal predictions
for the position or momentum of one, given (respectively) the position or
momentum of the other. EPR\ argued that this showed that quantum mechanics
must be incomplete, since measurement of the position (or momentum) of one
particle could not simultaneously give rise to a definite position (or
momentum) of the other particle, on pain of violation of locality. They
concluded that quantum mechanics, because it did not assign a position (or
momentum) to the other particle beforehand, must be incomplete.\footnote{The
argument of the EPR\ paper is notoriously convoluted, but I follow
\cite{Fin86} in regarding this as capturing Einstein's understanding of the
core argument.}

In 1964, John Bell proved a result based on David Bohm's streamlined version
of the EPR\ experiment \cite{Bell64}\cite{Boh51}. Instead of positions and
momenta, Bohm focuses on the spins of a pair of particles (either fermions or
bosons).\ Prepared in what has come to be known as a Bell\ state,
\begin{equation}
\psi=\frac{1}{\sqrt{2}}(\left\vert +x\right\rangle _{A}\left\vert
-x\right\rangle _{B}-\left\vert -x\right\rangle _{A}\left\vert +x\right\rangle
_{B})\text{,}%
\end{equation}
quantum mechanics predicts that a measurement of the component of spin of
particle $A$ in any direction (e.g., $\hat{z}$) is as likely to yield $+1$ as
$-1$ (in units of $\hbar/2$), and so the average value $\bar{A}$ is $0$.
However, quantum mechanics also indicates that an outcome of $+1$ for a
measurement of the spin of $A$ in the $\hat{z}$ direction is guaranteed to
yield an outcome of $-1$ for $B$ for a measurement of the spin of $B$ in the
$\hat{z}$ direction, etc. This is directly analogous to the correlations
between position and momentum measurements in the original EPR experiment.

In and of themselves, these phenomena offer no barrier to a hidden-variable
theory, since it is straightforward to explain such correlations by appealing
to a common cause -- the source -- and postulating that the particles emanate
from this source in (anti)correlated pairs. However, one must also account
for\ the way that the anticorrelation drops off as the\ angle between the
components of spin for the two particles increases (e.g., as $A$ rotates from
$\hat{x}$ toward $\hat{z}$ while $B$ remains oriented along the $\hat{x}$
direction). It was Bell's great insight to note that the quantum theory
implies that the anticorrelation is held onto more tightly than could be
accounted for by any \textquotedblleft local\textquotedblright\ theory ---
that is, any theory satisfying the seemingly natural condition known in the
literature variously as \emph{strong locality, Bell locality}, or
\emph{factorizability}. Bell showed that the predictions of a local theory
must satisfy a certain inequality, and that this inequality is violated by
quantum theory for appropriate choices of the components of spin to be
measured. Bell's result was widely understood to provide a barrier to the sort
of \textquotedblleft completion\textquotedblright\ of quantum mechanics
considered by Einstein. That is, Einstein's hope for a more fundamental theory
underlying quantum theory would have to violate \emph{strong locality}, of
which more below.

However, there is a further assumption known as the \emph{statistical
independence} assumption that is necessary for Bell's result. This assumption
is quite closely related to the assumption of local degrees of freedom, or the
absence of nonlocal constraints. Without the assumption, Bell's result does
not go through, and the possibility re-emerges of a local completion of
quantum theory after all.\footnote{A more detailed discussion of Bell's
derivation and the role of the Statistical Independence assumption can be
found in \cite{SW08a}.}

Rather than repeat the derivation of Bell's result, let me just focus on the
meaning of the two crucial assumptions of \emph{strong locality} and
\emph{statistical independence}. The physical situation we are attempting to
describe has the following form:%
\[%
%TCIMACRO{\FRAME{itbpF}{3.7204in}{0.7316in}{0in}{}{}{EPRboxnew.pdf}%
%width 3.7204in;  height 0.7316in;  depth 0in;  original-width 6.0528in;
%original-height 1.1537in;  cropleft "0";  croptop "1";  cropright "1";
%cropbottom "0";  filename 'EPRboxnew.pdf';file-properties "XNPEU";}}}%
%BeginExpansion
{\includegraphics[
natheight=1.153700in,
natwidth=6.052800in,
height=0.7316in,
width=3.7204in
]%
{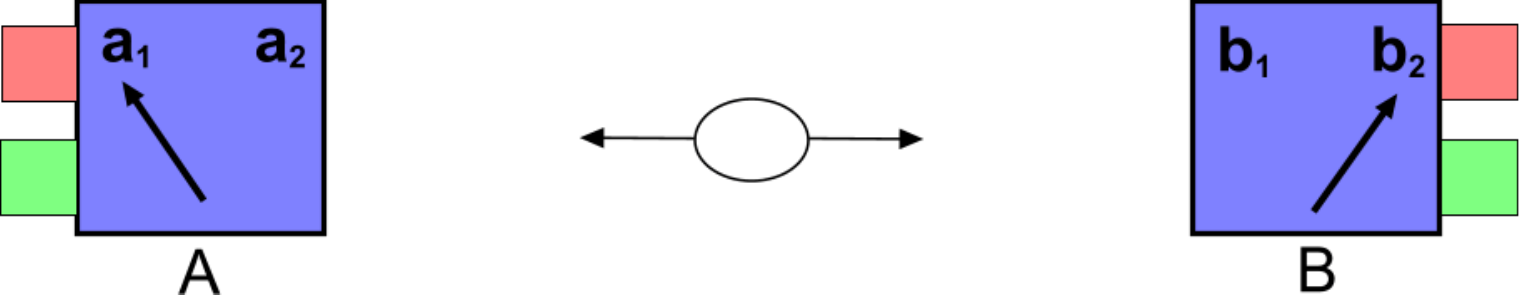}%
}%
%EndExpansion
\]
\noindent A source (represented by the ellipse) emits a pair of particles, or
in some other way causes detectors $A$ and $B$ to simultaneously (in some
reference frame) register one of two outcomes$.$ The detectors can be set in
one of two different ways, corresponding, in Bohm's version of the
EPR\ experiment, to a measurement of one of two different components of spin.

Let us now suppose that we have a theory that describes possible states of the
particles and which gives rise to either probabilistic or deterministic
predictions as to the results of various measurements one might make on the
particles. The state of the particle will be represented by a discrete or
continuous parameter $\lambda$, describing either a discrete set of states
$\lambda_{1},\lambda_{2}...$ or a continuous set. The expressions $\bar
{A}(a,\lambda)$ and $\bar{B}(b,\lambda)$ correspond to the expected (average)
values of measurements of properties $a$ and $b$ at detectors $A$ and $B$
(respectively) in a given state $\lambda$. (The appeal to \emph{average}
values allows for stochastic theories, in which a given $\lambda$ might give
rise to any number of different outcomes, with various probabilities.)

In general, one might suppose that $\bar{A}$ also depends on either the
detector setting $b$ or the particular outcome $B$ (i.e., $\bar{A}$ $=\bar
{A}(a,\lambda,b,B)$), and one might suppose the same for $\bar{B}$. That it
does not, that the expectation value $\bar{A}$ in a given state $\lambda$ does
\emph{not} depend on what one chooses to measure at B, or on the value of the
distant outcome $B$ (and vice-versa) is Bell's \emph{strong locality}
assumption. Given this assumption, one can write the expression $E(a,b,\lambda
)$ for the expected product of the outcomes of measurements of properties $a$
and $b$ in a given state $\lambda$ as
\begin{equation}
E(a,b,\lambda)=\bar{A}(a,\lambda)\bar{B}(b,\lambda)\text{.} \label{exp1}%
\end{equation}
This strong locality is also known as `factorizability', deriving as it does
from the fact that the joint probability of a pair of outcomes can be
factorized into the product of the marginal probabilities of each outcome. We
can thus represent the analysis of the experimental arrangement in this way,
where the expression for $E(a,b,\lambda)$ in the center encodes the assumption
of strong locality:%
\[%
%TCIMACRO{\FRAME{ihF}{4.6484in}{0.8622in}{0in}{}{}{EPRbox12new.pdf}%
%width 4.6484in;  height 0.8622in;  depth 0in;  original-width 7.5732in;
%original-height 1.3664in;  cropleft "0";  croptop "1";  cropright "1";
%cropbottom "0";  filename 'EPRbox12new.pdf';file-properties "XNPEU";}}}%
%BeginExpansion
{\includegraphics[
natheight=1.366400in,
natwidth=7.573200in,
height=0.8622in,
width=4.6484in
]%
{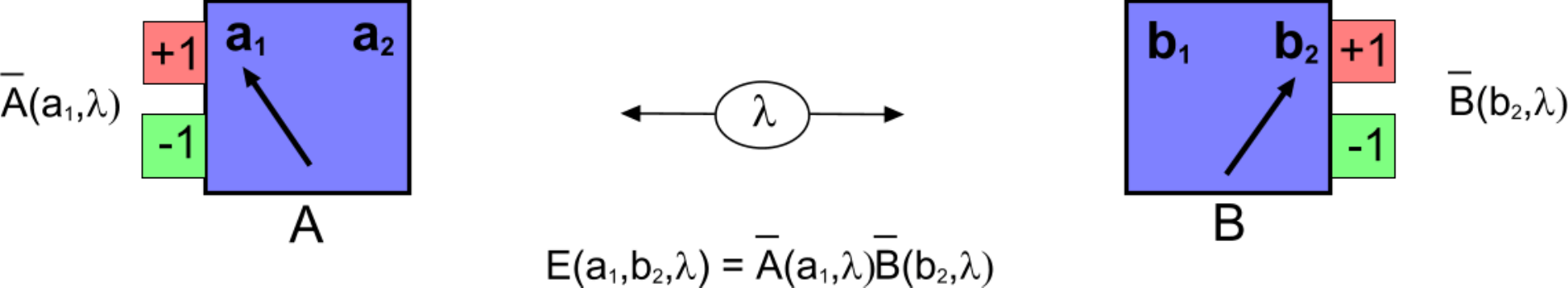}%
}%
%EndExpansion
\]

Now the further assumption required for Bell's result is that the probability
of a given state $\lambda$ is independent of the detector settings. In other
words, Bell assumes that the theory will be one in which
\begin{equation}
P(\lambda|a,b)=P(\lambda).\label{SI}%
\end{equation}
This is \emph{statistical independence. }For example, we might suppose that
the theory tells us that one of three states $\lambda_{1},\lambda_{2}%
,\lambda_{3}$ will be generated by our particle preparation procedure. This
condition tells us that the likelihood of obtaining any one of these states is
independent of how the detectors will be set at the time of detection. In
other words, knowledge of the future settings of those detectors (their
settings at the time the particles arrive) does not provide any further
information as to which of the three states was emitted.\footnote{Actually, a
slightly weaker condition than \emph{SI} is sufficient to derive the CHSH
inequality. \ See \cite{Fahm05} and the discussion thereof in section 3.3.1 of
\cite{Seev08}.}

The assumption of \emph{statistical independence }has been called into
question only infrequently, but when it has, the critique has often been
motivated by an appeal to the plausibility of Lorentz-invariant
\textquotedblleft backward causation\textquotedblright, whereby the change of
detector settings gives rise to effects that propagate along or within the
backward lightcone and thereby give rise to nontrivial initial correlations in
the particle properties encoded in $\lambda$ (e.g., \cite{Cos78}%
,\cite{Suth83},\cite{Pri96}). In my \cite{SW08a} I offer a critique of this
way of thinking. Here instead I would like to offer a rather different sort of
motivation for thinking that statistical independence might be violated,
coming as promised from the possibility of nonlocal constraints.

\begin{figure}[ptb]%
\centering
\includegraphics[
natheight=4.793700in,
natwidth=7.600000in,
height=2.5598in,
width=4.0421in
]%
{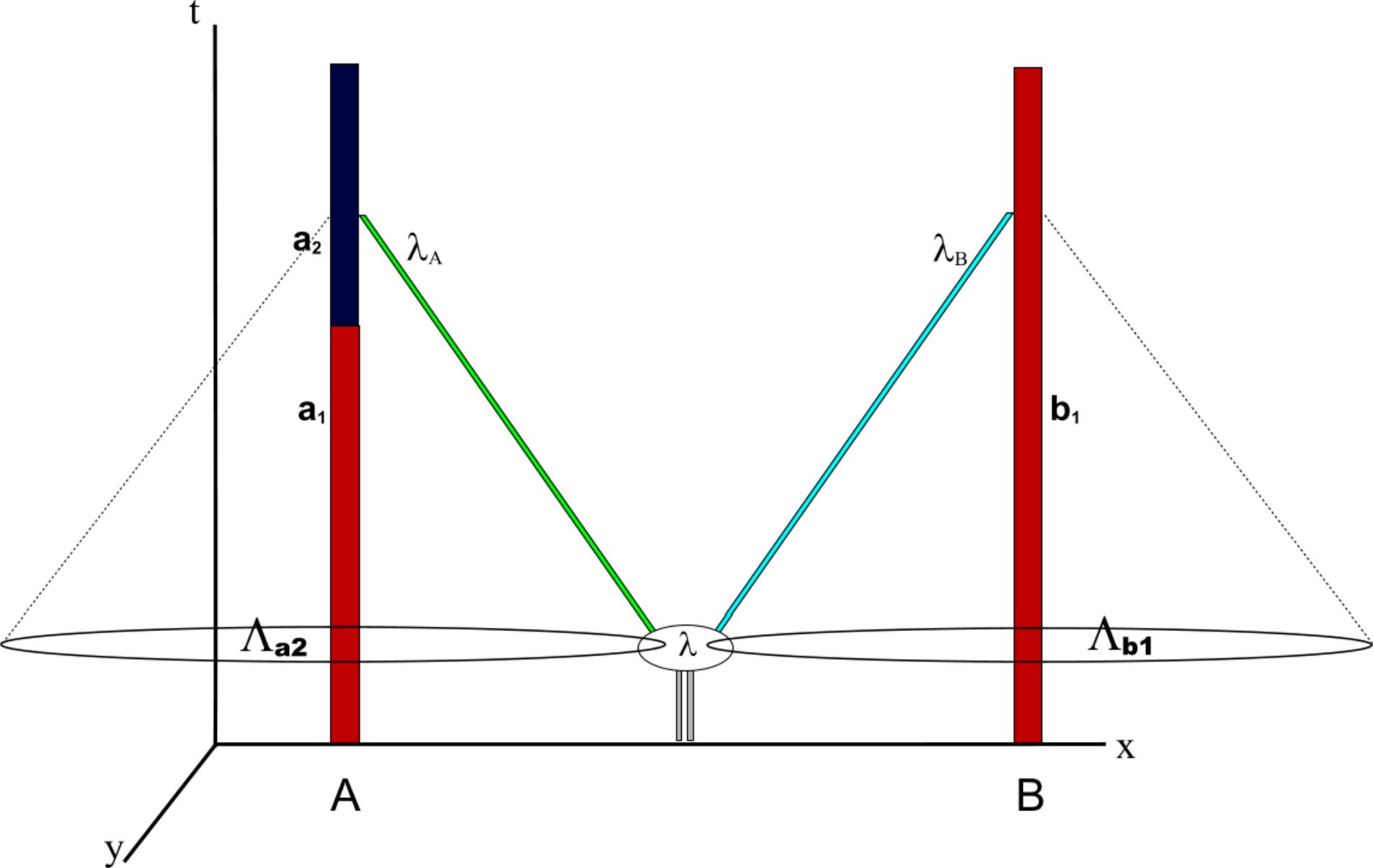}%
\caption{EPR:\ Spacetime diagram}%
\label{EPRfw}%
\end{figure}

Depicted in Figure \ref{EPRfw} is a run of the EPR-Bohm experiment in which
the setting of A is changed from $a_{1}$ to $a_{2}$ while the particles (or
whatever it is that emanates from the source) are in flight. What we have here
is a pair of particles traveling toward detectors $A$ and $B$, with detector
$A$ switching from setting $a_{1}$ to $a_{2}$ while the particles are in
flight, and detector $B$ simply set to $b_{1}$.%
%TCIMACRO{\FRAME{ftbpFU}{4.0421in}{2.5598in}{0pt}{\Qcb{EPR:\ Spacetime
%diagram}}{\Qlb{EPRfw}}{epr4.pdf}{\special{ language "Scientific Word";
%type "GRAPHIC";  maintain-aspect-ratio TRUE;  display "USEDEF";
%valid_file "F";  width 4.0421in;  height 2.5598in;  depth 0pt;
%original-width 7.6in;  original-height 4.7937in;  cropleft "0";  croptop "1";
%cropright "1";  cropbottom "0";  filename 'epr4.pdf';file-properties "XNPEU";}%
%}}%
%BeginExpansion
%EndExpansion

\noindent

Let's again suppose that the particles are in one of three states $\lambda
_{1},\lambda_{2},\lambda_{3}$. According to classical, relativistic physics,
the detector settings $a_{2}$ and $b_{1}$ are determined by the goings-on in
their past lightcones, which include the particle preparation event but also
far more.\ Suppose that setting $a_{2}$ is compatible with a variety of
initial data at the time of preparation, and the same for $b_{1}$. Let
$\Lambda_{a2}$ be the presumably large subset of microscopic states (in the
past lightcone of the detection event) consistent with a final detector
setting of $a_{2}$, and let $\Lambda_{b1}$ be those states compatible with
$b_{1}$. Though they reside in the past lightcone of the detection events, let
us suppose that the state of the particles, $\lambda_{1},\lambda_{2},$ or
$\lambda_{3}$, does not play a \emph{dynamical} role in determining the
setting of either detector. The question at hand is whether there is any
reason to think that, nevertheless, the state of the particles is
\emph{correlated} with the detector setting, which is to say whether the
theory constrains the state of the particles on the basis of $\Lambda_{a2}$
and $\Lambda_{b1}$.

Now, if the underlying theory is one in which local degrees of freedom are
independent, there is no reason to think that knowledge of $\Lambda_{a2}$ and
$\Lambda_{b1}$ should tell us anything about which of $\lambda_{1},\lambda
_{2},\lambda_{3}$ are realized. On the other hand, if there are nonlocal
constraints, then it may well be otherwise. Suppose that $\Lambda_{a2}$ is
compatible with $\lambda_{1}$ and $\lambda_{2}$ but incompatible with
$\lambda_{3}$. In other words, suppose that there are no microstates which
generate $a_{2}$ which are consistent with the particle pair starting in state
$\lambda_{3}$. Then we already have a violation of the statistical
independence condition, without even bothering yet to consider correlations
with the other detector $B$.

Of course, there may be, and typically are, many things going on in the past
lightcone of a detection event at the time the particle pair is produced. Most
of these will at least have the appearance of being irrelevant to the final
setting of the detector. There is certainly no guarantee that a nonlocal
constraint will generate the kind of correlations between detector settings
and specially prepared particles that we are talking about. The precise nature
of the nonlocal constraint or constraints that could explain quantum
correlations is a decidedly open question.

\section{Superdeterminism, conspiracy and free will}

The idea that the rejection of \emph{statistical independence} involves
preexisting and persisting\ correlations between subsystems has been broached
before, under the terms `conspiracy theory', `hyperdeterminism', and
`superdeterminism'. From here on, let us adopt 'superdeterministic' as a
generic term for this way of thinking about theories that violate this
condition. Bell \cite{Bel90b}, Shimony \cite{Shim85}, Lewis \cite{LewP06} and
others have suggested that superdeterministic theories imply some sort of
conspiracy on the part of nature. This is frequently accompanied by the charge
that the existence of such correlations is a threat to \textquotedblleft free
will\textquotedblright. Let me briefly address these worries before returning
to the big picture.

The idea that postulating a correlation between detector settings and particle
properties involves a \textquotedblleft conspiracy\textquotedblright\ on the
part of nature appears to derive from the idea that it amounts to postulating
that the initial conditions of nature have been set up by some cosmic
conspirator in anticipation of our measurements. It seems that the conspiracy
theorist is supposing that violations of \emph{statistical independence} are
not lawlike, but rather are \emph{ad hoc}. But nonlocal constraints are
lawlike, since (by definition), we require them to be consistent with the
dynamical evolution given by the laws of motion. If they exist, they exist at
every moment of time. This is no more a conspiracy than Gauss's law is a conspiracy.

Another worry about giving up \emph{statistical independence} and postulating
generic nonlocal, spacelike correlations has to do with a purported threat to
our \textquotedblleft free will\textquotedblright. This particular concern has
been the subject of renewed debate in the last couple of years, prompted in
part by an argument of Conway \&\ Kochen \cite{CK06}. The core of the worry is
that if detector settings are correlated with particle properties, this must
mean that we cannot \textquotedblleft freely choose\textquotedblright\ the
detector settings. However, as 't\ Hooft \cite{tHo07} points out, this worry
appears to be based on a conception of free will which is incompatible with
ordinary determinism, never mind superdeterminism. Hume \cite{Hume00} long ago
argued that such a conception of free will is highly problematic, in that it
is essential to the idea that we freely exercise our will that our thoughts
are instrumental in bringing about, which is to say determining, our actions.

\section{The Cosmos}

So much for the possible role of nonlocal constraints in underpinning quantum
phenomena. The other point of interest is early universe cosmology. Our
universe appears to have emanated from a big bang event around 14 billion
years ago, and to have been highly homogeneous for quite some time
thereafter.The cosmic microwave background radiation is a fossil remnant of
the time, around 400,000 years into the universe's existence, when radiation
effectively decoupled from matter, and this radiation appears to be quite
evenly distributed across the sky, with slight inhomogeneities which
presumably seeded later star and galaxy formation. 

The task of explaining the homogeneity of the early matter distribution is
known as the horizon problem. This, along with the flatness problem and
monopole problem, were for some time only explained by fine-tuning, which is
to say that they were not really explained at all.\ Later, inflationary models
entered the picture, and these provide a mechanism for generating
inhomogeneity in a more generic fashion. However, these models are still
speculative -- there is no direct evidence for an `inflaton' field -- and
moreover inflation itself requires rather special initial
conditions\cite{Pen89}.

The existence of a nonlocal constraint on the matter distribution and on the
state of the gravitational field might address one or more of these problems
without recourse to inflation. Certainly, a detailed description of the very
early universe requires few variables, since the universe looks essentially
the same from place to place with respect to both matter distribution (high
temperature, homogeneous) and spatial structure (flat).\ \ A reduction in the
number of variables is what we would expect from a constrained system, and any
constraint demanding that the matter distribution is identical from place to
place is indeed nonlocal. \ However, it is evidently not preserved under
dynamical evolution because of the action of gravity. One might speculate,
though, that the constraint holds between matter and gravitational degrees of
freedom, and that the early universe is simply a demonstration of one way to
satisfy it. \ The interplay of gravity and matter mix up the degrees of
freedom as time goes on, and the current remnant of these correlations are the
quantum correlations discussed above. \ 

\section{Conclusion}

The idea of using nonlocal constraints to account for the large-scale matter
distribution in the universe and the large-scale spacetime structure of the
universe is interesting but highly speculative, and the idea that these same
constraints might account for quantum correlations as well is even more
speculative. \ The most conservative strategy of exploration would be to
ignore cosmological scenarios and instead focus on the persistent and
experimentally repeatable correlations in the quantum realm. \ But I think it
is worth considering a connection between the two, if for no other reason than
the fact that it has proven difficult to construct a testable and sensible
quantum theory of gravity, suggesting that the relation between gravitation
and quantum phenomena might be different from anything heretofore explored.

A more conservative approach focusing just on quantum phenomena might ponder
the way in which the ordinarily superfluous gauge degrees of freedom of modern
gauge theories might serve as nonlocal hidden variables. The vector potential
in electrodynamics, for example, ordinarily plays no direct physical
role:\ only derivatives of the vector potential, which give rise to the
electric and magnetic fields, correspond to physical \textquotedblleft degrees
of freedom\textquotedblright\ in classical and quantum electrodynamics.\ The
Aharonov-Bohm effect shows that the vector potential does play an essential
role in the quantum theory, but the effect is still gauge-invariant. One might
nevertheless conjecture that there is an underlying theory in which the
potential \emph{does} play a physical role, one in which the physics is
\emph{not} invariant under gauge transformations. It may be impossible for us
to directly observe the vector potential, and the uncertainties associated
with quantum theory may arise from our ignorance as to its actual (and
nonlocally constrained) value. From this perspective, quantum theory would be
an effective theory which arises from \textquotedblleft modding
out\textquotedblright\ over the gauge transformations, with the so-called
gauge degrees of freedom being subject to a nonlocal constraint and accounting
for the correlations we observe in EPR-type experiments

I would conclude by reminding the reader that the sort of nonlocality under
discussion in no way violates either the letter or the spirit of relativity.
\ No influences travel faster than light. \ The idea is simply that there are
correlations between spatially separate degrees of freedom, and thus that the
fabric of nature is a more richly structured tapestry than we have heretofore believed.%

%TCIMACRO{\TeXButton{Endnotes}{\newpage\begingroup\renewcommand{\enotesize
%}{\normalsize}
%\setlength{\parskip}{2ex}
%\theendnotes\endgroup\newpage}}%
%BeginExpansion
\newpage\begingroup\renewcommand{\enotesize}{\normalsize}
\setlength{\parskip}{2ex}
\theendnotes\endgroup\newpage
%EndExpansion
\bibliographystyle{plain}
\bibliography{Aajour_x,caus,main}

\end{document}